\documentclass[12pt]{article}

\input{epsf.tex}

\newcommand{\E}{{\cal{E}}}

\renewcommand{\a}{\alpha}
\renewcommand{\l}{\lambda}

\newcommand{\be}{\begin{equation}}
\newcommand{\ee}{\end{equation}}
\newcommand{\bea}{\begin{eqnarray}}
\newcommand{\eea}{\end{eqnarray}}
\def\J#1#2#3#4{#1 {\it #2} {\bf #3} #4}

\def\PRL{\it Phys. Rev. Lett.}
\def\PRD{{\it Phys. Rev.} D}
\def\PR{\it Phys. Rev.}
\def\JETP{\it Sov. Phys. JETP}
\def\GRG{\it Gen. Relat. Grav.}

\def\APNY{\it Ann. Phys. (NY)}

\def\JMP{\it J. Math. Phys.}
\def\JPA{\it J. Phys. A}
\def\CQG{\it Class. Quantum Grav.}
\def\GC{\it Gravit. Cosmology}

\def\PLA{\it Phys. Lett. A}

\begin{document}
\title{How can exact and approximate solutions\\
of Einstein's field equations be compared?}
\author{V~S~Manko\dag\, and
E~Ruiz\ddag}
\date{}
\maketitle

\vspace{-1cm}

\begin{center}
\dag Departamento de F\'\i sica,\\ Centro de Investigaci\'on y de
Estudios Avanzados del IPN,\\ A.P. 14-740, 07000 M\'exico D.F.,
Mexico\\ \ddag Area de F\'\i sica Te\'orica, Universidad de
Salamanca,\\ 37008 Salamanca, Spain
\medskip
\end{center}

\vspace{.1cm}

\begin{abstract}
The problem of comparison of the stationary axisymmetric vacuum
solutions obtained within the framework of exact and approximate
approaches for the description of the same general relativistic
systems is considered. We suggest two ways of carrying out such
comparison: ($i$) through the calculation of the Ernst complex
potential associated with the approximate solution whose form on
the symmetry axis is subsequently used for the identification of
the exact solution possessing the same multipole structure, and
($ii$) the generation of approximate solutions from exact ones by
expanding the latter in series of powers of a small parameter. The
central result of our paper is the derivation of the correct
approximate analogues of the double--Kerr solution possessing the
physically meaningful {\it equilibrium configurations}. We also
show that the interpretation of an approximate solution originally
attributed to it on the basis of some general physical
suppositions may not coincide with its true nature established
with the aid of a more accurate technique.
\end{abstract}

\medskip


\vspace{1cm}


\newpage

\noindent{\bf 1. Introduction}

\vspace{.5cm}

\noindent Approximation methods had long proved to be an efficient
tool for treating specific general relativistic problems. For many
years their use was predominant since only a restricted number of
physically realistic exact solutions of Einstein's equations was
available before the development of modern solution generating
techniques [1--7]. But even now, when many exact solutions have
already been constructed, approximation methods are frequently
used for attacking some interesting problems not yet solved
exactly. Among the relatively recent successful applications of
the approximation schemes one can mention the analysis of the
frame--dragging effect in the field of a static charged massive
magnetic dipole \cite{Bon1} with the aid of the Martin--Pritchett
approximate solution \cite{MPr}, or a study of the equilibrium
problem of two static charged masses using a test--particle
approach \cite{Bon2}. The exact solutions which are in accord with
the above approximate findings were constructed later on
\cite{MMS,BMA}.

In spite of a rather peaceful coexistence of exact and
approximation approaches it appears that there does exist a
problem upon which the two approaches had produced some not fully
concurring results -- the vacuum equilibrium problem of two
spinning particles. There is only agreement of the exact results
with the well--known paper of Wald \cite{Wald1} in which he proved
within the framework of a test--particle approach that the
spin--spin interaction cannot balance the gravitational attraction
of two point--like particles. In his effort to develop a
far--reaching successive approximation procedure which would go
beyond the paper \cite{Wald1}, Bonnor derived an approximate
solution \cite{Bon3} which he considered applicable to any two
spinning uncharged particles, black holes or superextreme objects.
In that solution, however, the balance was not achievable too,
this time producing an apparent contradiction with the known exact
results. Indeed, it is well known that equilibrium states of two
superextreme particles do exist in the double--Kerr solution first
constructed by Kramer and Neugebauer \cite{KNe}. The general
analytic solution of the extended double--Kerr equilibrium problem
was given in \cite{MRS,MRu1},\footnote{Equilibrium states of two
identical superextreme Kerr constituents were first obtained and
analyzed by Dietz and Hoenselaers \cite{DHo}.} and in \cite{MRu2}
a simple relation was found between the coordinate distance at
which the balance occurs and the masses and angular momenta of the
balancing Kerr constituents.

This gives rise to an important question: how can one compare
exact and approximate solutions aimed at describing the same
physical situations? It is very important to have an answer to
this question for being able to interpret approximate solutions
correctly, establish limits of their validity, and give a recipe
for the rectification of an approximation scheme which for some
reasons has turned out to be incompatible with an exact model. In
this paper we shall develop an effective approach to the
comparison of the exact and approximate solutions based on the
following mutually complementary procedures: ($i$) identification
of the exact analogue of an approximate solution through the axis
expression of the Ernst potential \cite{Ernst} of the approximate
solution, and ($ii$) generation of the approximate solution from
an exact one by expanding the Ernst potential of the exact
solution in a series of powers of a small parameter with a
subsequent comparison of the result with the known approximate
solution. Whereas the procedure ($i$) provides one with a precise
interpretation of the approximate solution (which may not coincide
with the interpretation initially attributed to it), the procedure
($ii$) allows the construction of the approximate solution best
matching to the exact one. To illustrate our approach, we shall
apply it to the classical problem of two spinning particles,
taking as reference points the recent Bonnor's approximate
solution \cite{Bon3} and the exact double--Kerr solution in its
extended form \cite{MRu1}. Note that the solution \cite{Bon3} was
claimed to be appropriate for the approximate description of two
Kerr particles, but we will show that this affirmation is
erroneous.

The paper is organized as follows. In Section~2 we derive the
Ernst potential of the Bonnor solution and give the interpretation
of the latter by means of the corresponding exact solution. In
Section~3 we construct the approximate double--Kerr solution in
the subextreme case not admitting equilibrium configurations.
Section~4 deals with the approximate double--Kerr solutions
involving one or two superextreme particles for which the
corresponding equilibrium problems can be solved in full analogy
with the exact solution. In Section~5 we establish the relation of
the parameters entering approximate solutions to the individual
Komar masses and angular momenta of the particles. Section~6
presents a discussion of the results obtained. Concluding remarks
are contained in Section~7.

\newpage

\noindent{\bf 2. Interpretation of Bonnor's approximate solution}

\vspace{.5cm}

\noindent In this section we will show how a precise physical
interpretation of an approximate axisymmetric solution can be
established via an appropriately identified exact solution whose
properties can be ascribed to the approximate solution. Though the
existence of a solid basis for finding such an exact analogue may
look questionable at first sight, this is fortunately not the
case. The well--known Sibgatullin's integral method of the
construction of exact solutions \cite{Sib} permits one to generate
axisymmetric spacetimes from the axis expressions of their Ernst
complex potentials \cite{Ernst} which are used as the initial data
in the generation process. This means that all one needs for
having at hand an exact analogue of a known approximate solution
is to establish the form of the Ernst potential on the symmetry
axis of the latter; Sibgatullin's  method will then provide one
with the corresponding exact solution possessing the same axis
expression of the Ernst potential, and its properties can be
studied with the aid of the well established procedures.

This scheme can be used for establishing the physical
interpretation of the approximate vacuum solution for two spinning
particles \cite{Bon3}. Since this solution was obtained in a
somewhat special form, we first of all have to rewrite it using
the canonical axisymmetric line element\footnote{Throughout the
paper units are used in which Newton's gravitational constant $G$
and the speed of light $c$ are equal to unity.}
\be
d s^2=f^{-1}[e^{2\gamma}(d\rho^2+d z^2)+\rho^2d\varphi^2]-f(d
t-\omega d\varphi)^2, \label{Papa} \ee where the metric
coefficients $f$, $\gamma$ and $\omega$ depend on the coordinates
$\rho$, $z$ only; $\varphi$ and $t$ are the athimuthal angle and
time, respectively.

Observing that the line element (1) of \cite{Bon3} can be cast,
after changing the signature, into the form \bea d
s^2&=&f^{-1}[e^{\nu}(dr^2+d z^2)+r^2d\theta^2]-f(d
t-nf^{-1}d\theta)^2 \nonumber\\ &-&f^{-1}(r^2-n^2-lf)d\theta^2,
\label{line_B} \eea we obtain formulae relating notations of the
paper \cite{Bon3} and ours: \be r=\rho, \quad z=z, \quad
\nu=2\gamma, \quad n=f\omega, \quad r^2-n^2-lf=0. \label{rel} \ee

Now, bearing in mind (\ref{rel}) and using formulae (14)--(20) of
Ref.~\cite{Bon3}, we are able to rewrite the Bonnor solution in
its standard form: \bea f&=&1-2\l\left(\frac{m_1}{R_1}
+\frac{m_2}{R_2}\right)+2\l^2\left[\left(\frac{m_1}{R_1}
+\frac{m_2}{R_2}\right)^2\right. \nonumber\\
&-&\left.\left(\frac{j_1(z-b)}{R_1^3}
+\frac{j_2(z+b)}{R_2^3}\right)^2\right], \nonumber\\
\gamma&=&\frac{1}{2}\l^2\left(\frac{3j_1j_2-2b^2m_1m_2}{2b^4}
-\frac{m_1^2\rho^2}{R_1^4}-\frac{m_2^2\rho^2}{R_2^4}
+\frac{m_1m_2(\rho^2+z^2-b^2)}{b^2R_1R_2} \right. \nonumber\\
&-&\frac{j_1j_2[3(\rho^2+z^2-b^2)^3+2b^2\rho^2(9\rho^2+9z^2-b^2)]}
{2b^4R_1^3R_2^3} \nonumber\\ &+&\left.\frac{j_1^2\rho^2[\rho^2
-8(z-b)^2]}{2R_1^8}+\frac{j_2^2\rho^2[\rho^2
-8(z+b)^2]}{2R_2^8}\right), \nonumber\\
\omega&=&-2\l\rho^2\left(\frac{j_1}{R_1^3}+\frac{j_2}{R_2^3}\right)
+\l^2\left[\frac{m_1j_2+m_2j_1}{b^2}-2\rho^2
\left(\frac{m_1j_1}{R_1^4}+\frac{m_2j_2}{R_2^4}\right) \right.
\nonumber\\
&-&\frac{m_1j_2[\rho^4+2\rho^2(z^2+bz+2b^2)+(z+b)^2(z^2-b^2)]}
{b^2R_1R_2^3} \nonumber\\ &-& \left.
\frac{m_2j_1[\rho^4+2\rho^2(z^2-bz+2b^2)+(z-b)^2(z^2-b^2)]}
{b^2R_1^3R_2}\right], \nonumber\\ R_1&=&\sqrt{\rho^2+(z-b)^2},
\quad R_2=\sqrt{\rho^2+(z+b)^2}. \label{sol_B} \eea In the above
formulae we have left Bonnor's notations for the masses $m_1$,
$m_2$ but instead of his $h_1$, $h_2$ we have used ours
$j_1=-h_1$, $j_2=-h_2$ as angular momenta of the particles; the
parameter $b$ defines location of the particles on the symmetry
axis (the points $z=\pm b$). We have introduced an auxiliary
nondimensional small parameter $\l$ by the formulae
\bea&&f=f^{(0)}+\l f^{(1)}+\l^2 f^{(2)}, \quad \nu=\nu^{(0)}+\l
\nu^{(1)}+\l^2 \nu^{(2)}, \nonumber\\ &&n=n^{(0)}+\l n^{(1)}+\l^2
n^{(2)} \eea ($f^{(i)}$, $\nu^{(i)}$, $n^{(i)}$ are defined in
\cite{Bon3}, index $i$ denoting the order of approximation), and
it helps to control the order of approximation in Bonnor's scheme
(one can always put $\l=1$ at the end). Note that the metric
functions (\ref{sol_B}) cannot contain terms with $\l^n$, $n>2$,
because the solution is given in the second approximation. Mention
also that the masses $m_i$ and angular momenta $j_i$ are
quantities of the same (first) order in $\l$.

The metric coefficients $f$, $\gamma$, $\omega$ defined by
(\ref{sol_B}) are approximate solutions to Einstein's equations in
the stationary axisymmetric case (see, e.g., Section~7.1 of
\cite{Wald2}) \bea&&f\Delta f=(\vec\nabla
f)^2-\rho^{-2}f^4(\vec\nabla\omega)^2, \nonumber\\
&&\vec\nabla(\rho^{-2}f^2\vec\nabla\omega)=0, \nonumber\\
&&\gamma_{,\rho}=\frac{1}{4}\rho f^{-2}[f_{,\rho}^2-f_{,z}^2
-\rho^{-2}f^4(\omega_{,\rho}^2-\omega_{,z}^2)], \nonumber\\
&&\gamma_{,z}=\frac{1}{2}\rho f^{-2}(f_{,\rho}f_{,z}
-\rho^{-2}f^4\omega_{,\rho}\omega_{,z}^2), \label{eq_E} \eea where
a comma in subindices denotes partial differentiation, and the
operators $\Delta$, $\vec\nabla$ have the form
\be
\Delta=\frac{\partial^2}{\partial\rho^2}+\frac{1}{\rho}
\frac{\partial}{\partial\rho}+\frac{\partial^2}{\partial z^2},
\quad \vec\nabla=\vec\rho_0\frac{\partial}{\partial\rho} +\vec
z_0\frac{\partial}{\partial z} \ee ($\vec\rho_0$ and $\vec z_0$
are unit vectors).

The first two equations in (\ref{eq_E}) can be rewritten as a
single Ernst equation for a complex potential $\E$ \cite{Ernst}
\be
(\E+\bar\E)\Delta\E=2(\vec\nabla\E)^2, \label{Ernst} \ee a bar
over a symbol denoting complex conjugation.

The relation of $\E$ to the functions $f$ and $\omega$ is the
following:
\be
\E=f+i\Omega, \quad \Omega_{,\rho}= \rho^{-1}f^2\omega_{,z}, \quad
\Omega_{,z}= -\rho^{-1}f^2\omega_{,\rho}. \label{E} \ee

Sibgatullin's method \cite{Sib} provides a straightforward
procedure for the construction of exact solutions of equation
(\ref{Ernst}) corresponding to a given function
$e(z)=\E(\rho=0,z)$ representing the axis value of the Ernst
potential $\E$. Therefore, in order to identify the exact analogue
of Bonnor's approximate solution (\ref{sol_B}), we have to find
from (\ref{sol_B}) and (\ref{E}) the axis value $e_B(z)$ of the
corresponding Ernst potential in the second approximation. Having
obtained $e_B(z)$, we shall be able to point out what exact
solution corresponds to this function and, besides, to use
$e_B(z)$ for the calculation of the multipole moments of both the
approximate and exact solutions with the aid of the Fodor {\it et
al.} procedure \cite{FHP}.

The integration of equations (\ref{E}) for $\Omega$, with $f$ and
$\omega$ defined in (\ref{sol_B}), can be readily performed using
the Mathematica computer programme \cite{Wol}. The resulting
expression, multiplied by $i$, should then be added to $f$,
yielding the desired Ernst potential of the Bonnor solution \bea
&&\E_B=1-2\l M+2\l^2(M^2-R^2)-2i\l R \nonumber\\ &&+4i\l^2\left(
\frac{ z(m_1j_2+m_2j_1)(\rho^2+z^2-b^2)+
b(m_1j_2-m_2j_1)(\rho^2-z^2+b^2) }{R_1^3R_2^3}\right.\nonumber\\
&&+\left.\frac{m_1j_1(z-b)}{R_1^4}
+\frac{m_2j_2(z+b)}{R_2^4}\right), \nonumber\\
&&M=\frac{m_1}{R_1}+\frac{m_2}{R_2}, \quad
R=\frac{j_1(z-b)}{R_1^3}+\frac{j_2(z+b)}{R_2^3}. \eea

On the upper part of the symmetry axis, $\rho=0$, $z>b$, the
potential obtained takes the form \bea e_B(z)&=&\E_B(\rho=0,z)
=1-\frac{2\l N}{(z^2-b^2)^2}+\frac{2\l^2 N^2}{(z^2-b^2)^4},
\nonumber\\ N&=&(z^2-b^2)[z(m_1+m_2)+b(m_1-m_2)] \nonumber\\
&+&i(j_1+j_2)(z^2+b^2)+2ibz(j_1-j_2). \label{axis_B} \eea

Expression (\ref{axis_B}) is the ratio of two polynomials of the
eighth order in $z$, so the exact solution corresponding to this
axis data necessarily falls into the family of $2N$--soliton
solutions discussed in detail in \cite{MRu3}. Moreover, as the
denominator of $e_B(z)$ is the real function of $z$, it should be
a special member of the $N=4$ subfamily of the soliton solutions
which in the general case involves 16 real parameters, contrary to
only 5 parameters of the axis data (\ref{axis_B}).

To see what physical situation describes the exact solution whose
Ernst potential on the symmetry axis has the form (\ref{axis_B}),
one has to recall that after the application of Sibgatullin's
method to the data (\ref{axis_B}) the resulting solution will
contain the functions $r_i=\sqrt{\rho^2+(z-\a_i)^2}$, where $\a_i$
are roots of the algebraic equation
\be
e_B(z)+\bar e_B(z)=0. \label{al_eqB} \ee Apparently, this equation
has 8 roots which can assume real values or occur in complex
conjugate pairs, hence the solution will describe special
configurations of 4 spinning objects because each pair of $\a_i$
determines one sub- or one superextreme constituent.

For a more precise analysis we need the explicit form of $\a_i$.
Obviously, equation (\ref{al_eqB}) cannot be solved exactly, but
its approximate roots up to the second order in $\l$ are
available. These should be searched for in the form $\pm
b+z_1\l^{1/2}+z_2\l+z_3\l^{3/2}+z_4\l^2$, the constant
coefficients $z_1$, $z_2$, $z_3$, $z_4$ to be found from
(\ref{al_eqB}). Then one gets the following 8 roots of equation
(\ref{al_eqB}): \bea
&&\a_1=b+\sqrt{j_1}\l^{1/2}+A_1\l^{3/2}+A_{12}\l^2, \nonumber\\
&&\a_2=b-i\sqrt{j_1}\l^{1/2}+iA_1\l^{3/2}+A_{12}\l^2, \nonumber\\
&&\a_3=b+i\sqrt{j_1}\l^{1/2}-iA_1\l^{3/2}+A_{12}\l^2, \nonumber\\
&&\a_4=b-\sqrt{j_1}\l^{1/2}-A_1\l^{3/2}+A_{12}\l^2, \nonumber\\
&&\a_5=-b+\sqrt{j_2}\l^{1/2}+A_2\l^{3/2}-A_{12}\l^2, \nonumber\\
&&\a_6=-b-i\sqrt{j_2}\l^{1/2}+iA_2\l^{3/2}-A_{12}\l^2, \nonumber\\
&&\a_7=-b+i\sqrt{j_2}\l^{1/2}-iA_2\l^{3/2}-A_{12}\l^2,
\label{alfas_B}
\\ &&\a_8=-b-\sqrt{j_2}\l^{1/2}-A_2\l^{3/2}-A_{12}\l^2,
\nonumber\\ &&A_1=\frac{2b^2m_1^2+j_1j_2}{8b^2\sqrt{j_1}}, \,
A_2=\frac{2b^2m_2^2+j_1j_2}{8b^2\sqrt{j_2}}, \,
A_{12}=\frac{2b^2m_1m_2-j_1j_2}{8b^3}, \nonumber \eea which means
that independently of the sign of $j_1$ and $j_2$ these $\a$s
determine a system formed by two sub- and two superextreme
constituents. More precisely, we have a system of two separated
{\it compound} objects, each of which is composed of the
overlapping sub- and superextreme constituents. One easily arrives
at this conclusion by analyzing for instance the expressions for
$\a_1$, $\a_2$, $\a_3$, $\a_4$, of which $\a_1$ and $\a_4$ define
a subextreme constituent (assuming $j_1>0$), whereas $\a_2$ and
$\a_3$ define a superextreme constituent; these two overlapping
constituents form a compound spinning object. Another compound
object is formed by a sub- and a superextreme constituents defined
by the constants $\a_5$, $\a_6$, $\a_7$, $\a_8$ (see Fig.~1).

Therefore, the physical interpretation of the Bonnor solution
(\ref{sol_B}) supplied by the corresponding exact solution is the
following: {\it it represents the second approximation to a
special case of the $N=4$ subfamily of multi--soliton exact
solutions describing two compound objects located on the symmetry
axis and kept apart by a strut in between}. Such systems of two
compound objects have been considered in the paper \cite{GRG}.

It should be also pointed out that Bonnor's solution cannot be
considered as an approximation to the well--known double--Kerr
spacetime \cite{KNe}. Although the general $N=4$ soliton metric
(the quadruple--Kerr solution) possesses 16 arbitrary real
parameters and permits in principle the reduction to the simpler
$N=2$ double--Kerr case, such a reduction is not possible for
$\a$s defined in (\ref{alfas_B}) because of the insufficient
number of constants required for performing the desired reduction.
Correct approximations to the double--Kerr solution will be
obtained in the next two sections.

To conclude this part, below we write down the first five complex
multipole moments of the Bonnor solution calculated from
(\ref{axis_B}) with the aid of the Fodor {\it et al.} procedure
\cite{FHP}: \bea &&P_0=\l(m_1+m_2), \quad
P_1=\l[b(m_1-m_2)+i(j_1+j_2)], \nonumber\\
&&P_2=b\l[b(m_1+m_2)+2i(j_1-j_2)], \nonumber\\
&&P_3=b^2\l[b(m_1-m_2)+3i(j_1+j_2)], \nonumber\\
&&P_4=b^3\l[b(m_1+m_2)+4i(j_1-j_2)]. \eea

The form of these multipoles confirms the interpretation of $m_1$,
$m_2$ as masses, and $j_1$, $j_2$ as angular momenta. However, a
more subtle analysis is needed to see whether $m_i$ and $j_i$
coincide with the individual Komar masses and angular momenta
\cite{Kom} of the particles (see our Section~5).

\vspace{.5cm}

\noindent{\bf 3. Approximate double--Kerr solution in the
black--hole case}

\vspace{.5cm}

\noindent When a problem has an exact solution of a complicated
form, one may think about the conversion of that solution into an
approximate one under some simplifying assumptions. The generation
of approximate solutions from the already known exact solutions
has obvious advantages compared to the development of
approximation schemes exclusively on the basis of some general
suppositions and physical intuition, first of all because the
approximation in the former case will not contradict the existing
exact results but will only permit their presentation in a simpler
form. On the other hand, there is always a danger that an
approximate method not orientated at the exact solution will lead
to a result contradicting the exact findings and, besides, will be
unable to identify the origin of the contradiction.

Usually the exact solutions constructed with the aid of the modern
solution generating techniques, especially those describing the
many--body configurations, involve the parameter sets most
adjusted to the mathematical structure of the solutions but which
may look ``unphysical'' to non--experts of the generation methods.
At the same time, a rather sophisticated relation of the
parameters of the generated solutions to the physical
characteristics of spacetimes they describe certainly may course
difficulties for introducing the simplifying physical assumptions
for a possible approximate analysis of the generated solutions.

The aim of this section and of the next one is to demonstrate that
within the framework of Sibgatullin's method it is possible not
only to construct exact solutions but also develop a consistent
approach to the analysis of their physical properties on the basis
of reliable approximation schemes. Below we shall obtain
approximate analogues of the well--known double--Kerr solution
\cite{KNe} making use of the unique opportunities offered by
Sibgatullin's method -- the possibility of the unified treatment
of the sub- and superextreme cases of spinning particles and a
very clear relations of all the parameters of the solution to the
axis data and, consequently, to the relativistic multipole
moments. The Ernst potential of the extended double--Kerr solution
has the form \cite{MRu1} \bea
\E&=&\frac{\Lambda+\Gamma}{\Lambda-\Gamma}, \quad
\Lambda=\sum_{1\leq i<j\leq4}\lambda_{ij}r_ir_j, \quad
\Gamma=\sum_{i=1}^4\nu_ir_i,\nonumber\\
\lambda_{ij}&=&(-1)^{i+j}(\a_i-\a_j)(\a_{i'}-\a_{j'})X_iX_j, \quad
(i',j'\neq i,j;\,\, i'<j'), \nonumber\\
\nu_{i}&=&(-1)^{i}(\a_{i'}-\a_{j'})(\a_{i'}-\a_{k'})
(\a_{j'}-\a_{k'})X_i, \nonumber\\ &&(i',j',k'\neq i;\,\, i'<j'<k')
\nonumber\\ X_i&=&\frac{(\a_i-\bar\beta_1)(\a_i-\bar\beta_2)}
{(\a_i-\beta_1)(\a_i-\beta_2)}, \quad
r_i=\sqrt{\rho^2+(z-\a_i)^2}, \label{DK} \eea The parameters
entering (\ref{DK}) are $\beta_1$, $\beta_2$ which can assume
arbitrary complex values, and $\a_i$, $i=1,2,3,4$, which can take
arbitrary real values or occur in complex conjugate pairs (without
loss of generality we can assign to $\a_i$ the following order:
${\rm Re}\a_1\ge{\rm Re}\a_2\ge{\rm Re}\a_3\ge{\rm Re}\a_4$). On
the symmetry axis $\E$ is defined by the expression
\be
e(z)=1+\frac{e_1}{z-\beta_1}+\frac{e_2}{z-\beta_2},
\label{axis_DK} \ee where $e_1$ and $e_2$ are two arbitrary
complex constants.

Note that $\a_i$ are roots of Sibgatullin's algebraic condition
\be
e(z)+\bar e(z)=0, \label{alg_cond} \ee which means that the
following relation holds:
\be
2+\sum\limits_{l=1}^2\left(\frac{e_l}{z-\beta_l} +\frac{\bar
e_l}{z-\bar\beta_l}\right)=\frac{2\prod_{n=1}^4(z-\a_n)}
{\prod_{k=1}^2(z-\beta_k)(z-\bar\beta_k)}, \ee whence one obtains
the connection between the parameters $e_l$ and $\a_n$
\be
e_1=\frac{2\prod_{n=1}^4(\beta_1-\a_n)}
{(\beta_1-\beta_2)(\beta_1-\bar\beta_1)(\beta_1-\bar\beta_2)},
\quad e_2=\frac{2\prod_{n=1}^4(\beta_2-\a_n)}
{(\beta_2-\beta_1)(\beta_2-\bar\beta_1)(\beta_2-\bar\beta_2)}. \ee

In the papers \cite{MRu1,MRu2} we presented the general solution
to the equilibrium problem of two Kerr particles and found a
simple relation between the coordinate distance at which
equilibrium takes place and individual Komar masses and angular
momenta of the balancing constituents. The question which now
would be of interest to answer in view of the paper \cite{Bon3} is
whether the double--Kerr equilibrium problem can be solved with
the aid of some approximation procedure? Below we are going to
show that the answer to this question is `yes' and it is supported
by an approximation scheme which is able to reproduce the exact
results.

In order to develop the desired physically meaningful
approximation procedure for the double--Kerr solution it is likely
first of all to rewrite the formulae (\ref{DK}) in a more
``physical'' parameter set which would be mathematically
equivalent to the parameter set used in (\ref{DK}). The
reparametrization of the axis data (\ref{axis_DK}) can be done in
analogy with a single Kerr solution \cite{Kerr} whose axis
expression of the potential $\xi=(1-\E)/(1+\E)$ has the
form\footnote{The potential $\xi=(\E-1)/(\E+1)$ was first
introduced by Ernst in his famous paper \cite{Ernst}. Changing the
sign in the original expression is explained by the successful use
of the modified potential in the Fodor {\it et al.} procedure
\cite{FHP} for the calculation of multipole moments.}
\be
x(z)=\frac{1-e(z)}{1+e(z)}=\frac{m}{z-ia}, \label{axis_Kerr} \ee
where $m$ is the total mass and $a$ is the angular momentum per
unit mass of the Kerr source.

In the case of the double--Kerr solution we simply generalize
(\ref{axis_Kerr}) to the expression
\be
x(z)=\frac{m_1+i\nu}{z-b-ia_1}+\frac{m_2-i\nu}{z+b-ia_2},
\label{axis_xi} \ee taking $m_1$, $m_2$ as the ``masses'' of the
Kerr particles, and $a_1$, $a_2$ as their ``angular momenta per
unit mass''; the parameter $b$ denotes displacements of the
particles from the origin of coordinates along the $z$--axis, and
$\nu$ represents the angular momentum of the part of the symmetry
axis separating the particles.

From (\ref{axis_xi}) we obtain the corresponding expression of the
Ernst potential $\E$ \bea e(z)&=&\frac{1-x(z)}{1+x(z)}=
\frac{e_-}{e_+}, \nonumber\\ e_\pm&=&(z-b-ia_1)(z+b-ia_2)
\nonumber\\ &\pm&(m_1+i\nu)(z+b-ia_2)\pm(m_2-i\nu)(z-b-ia_1).
\label{axis_DK2} \eea

From (\ref{axis_xi}) the expressions of the total mass $M$ and
total angular momentum $J$ of the system can be readily found with
the aid of the Fodor {\it et al.} procedure:
\be
M=m_1+m_2, \quad J=m_1a_1+m_2a_2+2b\nu. \ee This supports our
physical interpretation of the parameters, even though their
precise relation to the individual Komar masses and angular
momenta will be established later on in Section~5. It is worth
pointing out that the parametrization (\ref{axis_DK2}) of the
Ernst potential involves 6 arbitrary real parameters and it has
one advantage compared to the axis expression (\ref{axis_DK})
which is the absence of the unphysical NUT parameter that
guarantees the asymptotic flatness of the solution from the very
beginning. Up to this NUT parameter, the two parametrizations are
totally equivalent in a sense that both of them represent {\it
all} possible stationary asymptotically flat configurations of two
Kerr particles.

Now it is possible to develop an approximation scheme as follows.
We can assume the values of the parameters $m_1$, $m_2$, $a_1$,
$a_2$, $\nu$ to be small compared to the values of $b$, which is
equivalent to considering them proportional to some auxiliary
non--dimensional small parameter $\l$. Then an approximate
expression of the potential $\E$ can be found as an expansion in
powers of $\l$ up to the desired order. Technically this consists
in ($a$) finding approximate expressions for $\a_n$ from equation
(\ref{alg_cond}), ($b$) identifying the expansions for $X_n$ from
the formulae (\ref{DK}), (\ref{axis_DK2}) in which the approximate
expressions for $\a_n$ should be used, ($c$) obtaining the
expansions for $r_n$ corresponding to approximate $\a$s, and
finally ($d$) the expansion of the potential $\E$ in powers of
$\l$ after the substitution of $\a_n$, $X_n$, $r_n$ previously
obtained into the formulae (\ref{DK}). Once the potential $\E$ is
found, the approximate expressions for the metric coefficients
$\omega$ and $\gamma$ can be derived from equations (\ref{E}) and
(\ref{eq_E}) (note that in spite of the availability of the exact
expressions for $\omega$ and $\gamma$ \cite{MRS} it looks more
attractive to carry out additional integration instead of working
out the expansions of the known formulae).

We have to distinguish between the following three major
approximation schemes which must be treated separately.

\underline{\it Scheme~1.} All five parameters $m_1$, $m_2$, $a_1$,
$a_2$, $\nu$ are assumed to be small compared to $b$. In this
case, making the substitutions $m_i\to\l m_i$, $a_i\to\l a_i$,
$\nu\to\l\nu$ in (\ref{axis_DK2}), we see that Sibgatullin's
condition (\ref{alg_cond}) at the zeroth order in $\l$ yields the
equation
\be
(z^2-b^2)^2=0, \ee whose double roots $+b$ and $-b$ should be
taken as the zeroth approximation for the parameters $\a_i$:
$\a_1=\a_2=b$, $\a_3=\a_4=-b$ (see Fig.~2a).

\underline{\it Scheme~2.} In this scheme, $m_1$, $m_2$, $\nu$ are
small parameters compared to $b$, while $a_1$ and $a_2$ are of the
zeroth order in $\l$, i.e., `comparable' with $b$. After the
substitutions $m_i\to\l m_i$, $\nu\to\l\nu$, the zeroth order in
$\l$ of equation (\ref{alg_cond}) takes the form
\be
(z^2-2bz+b^2+a_1^2)(z^2+2bz+b^2+a_2^2)=0, \label{Sib0_2} \ee and
the roots of this equation provide us with the zeroth
approximation of $\a_i$: $\a_1=b-ia_1$, $\a_2=b+ia_1$,
$\a_3=-b-ia_2$, $\a_4=-b+ia_2$ (Fig.~2b).

\underline{\it Scheme~3.} In this scheme we assume that only one
of the constants $a_1$, $a_2$ is comparable with $b$, say, $a_2$,
whereas $m_1$, $m_2$, $a_1$, $\nu$ are small parameters. Equation
(\ref{alg_cond}) in this case leads to
\be
(z-b)^2(z^2+2bz+b^2+a_2^2)=0, \ee giving the following zero-order
approximations for $\a_i$: $\a_1=\a_2=b$, $\a_3=-b-ia_2$,
$\a_4=-b+ia_2$ (Fig.~2c).

Remarkably, the schemes~1--3 correspond to the configurations
composed of two subextreme constituents, of two superextreme
constituents, and of one sub- and one superextreme constituents of
the extended exact double--Kerr solution, respectively (we remind
that in the exact solution a pair of real--valued $\a$s defines a
subextreme constituent, and a pair of complex conjugate $\a$s
defines a superextreme Kerr constituent). Mention also that only
the first scheme can be considered as a sort of a
``point--like--particles approximation'' since, as will be seen
later on, it involves two real distances
$R_1=\sqrt{\rho^2+(z-b)^2}$ and $R_2=\sqrt{\rho^2+(z+b)^2}$,
whereas the remaining schemes involve, respectively, four and
three different square roots $R_i$ due to the presence of two
(scheme~2) and one (scheme~3) superextreme objects whose absolute
values of the angular momenta per unit mass exceed the individual
masses.

Below we shall work out in detail the approximation scheme~1;
schemes~2 and 3 will be considered in the next section.

\underline{\it Approximate double--Kerr solution via scheme~1.}
Without loss of generality we can set $b=1$ in the previous
formulae (this simply means the rescaling of the parameters and
coordinates) for simplifying a bit the expressions we are going to
obtain. Like the paper \cite{Bon3}, we shall be interested in the
expansion of $\E$ and corresponding metric functions in powers of
$\l$ up to $\l^2$ (the second approximation). Then we have to
calculate $\a_i$ up to $\l^3$ for obtaining the correct
expressions for $X_i$ in the second order of $\l$. After
introducing the factor $\l$ into the parameters $m_i$, $a_i$,
$\nu$, equation (\ref{alg_cond}) takes the form \bea
(z^2-1)^2&-&\l^2[(m_1^2-a_1^2)(z+1)^2+(m_2^2-a_2^2)(z-1)^2
\nonumber\\ &+&2m_1m_2(z^2-1)+4\nu^2]+2\nu\l^3(m_1+m_2)[a_1(z+1)
\nonumber\\
&-&a_2(z-1)]-\l^4[m_1^2a_2^2+m_2^2a_1^2+\nu^2(a_1-a_2)^2
\nonumber\\ &+&a_1a_2(2m_1m_2-a_1a_2)]=0. \label{sib1} \eea

The roots of this equation can be searched for in the form
$\pm1+z_1\l+z_2\l^2+z_3\l^3$, the coefficients $z_i$ to be
determined from (\ref{sib1}). The resulting expressions for $\a_i$
are the following (in view of the cumbersome coefficients $z_3$ we
write out $\a$s only up to the terms $\l^2$, but it should be
remembered that the coefficients $z_3$ cannot be omitted during
the calculations): \bea \a_1=1+\l\sqrt{m_1^2+\nu^2-a_1^2}
+\frac{\l^2}{2}\left(m_1m_2-\nu^2-\frac{a_1\nu(m_1+m_2)}
{\sqrt{m_1^2+\nu^2-a_1^2}}\right), \nonumber\\
\a_2=1-\l\sqrt{m_1^2+\nu^2-a_1^2}
+\frac{\l^2}{2}\left(m_1m_2-\nu^2+\frac{a_1\nu(m_1+m_2)}
{\sqrt{m_1^2+\nu^2-a_1^2}}\right), \nonumber\\
\a_3=-1+\l\sqrt{m_2^2+\nu^2-a_2^2}
-\frac{\l^2}{2}\left(m_1m_2-\nu^2+\frac{a_2\nu(m_1+m_2)}
{\sqrt{m_2^2+\nu^2-a_2^2}}\right), \nonumber\\
\a_4=-1-\l\sqrt{m_2^2+\nu^2-a_2^2}
-\frac{\l^2}{2}\left(m_1m_2-\nu^2-\frac{a_2\nu(m_1+m_2)}
{\sqrt{m_2^2+\nu^2-a_2^2}}\right). \label{alf1} \eea

The expansion of the constant objects $X_i$ from (\ref{DK})
associated with the parameters $\a_i$ can be most conveniently
found with the aid of the formula
\be
X_i=\left.\frac{\bar e_+}{e_+}\right|_{z=\a_i}, \ee where $e_+$ is
defined in (\ref{axis_DK2}).

The expansion of $X_i$ should be performed up to the second power
of $\l$; this yields (to avoid the complicated expressions, we
give explicitly only the terms linear in $\l$) \bea
X_1&=&\frac{\sqrt{m_1^2+\nu^2-a_1^2}+ia_1}{m_1+i\nu}
-\frac{\l}{2(m_1+i\nu)}\Biggl(2a_1a_2-i\nu(m_1+m_2) \nonumber\\
&&-2ia_2\sqrt{m_1^2+\nu^2-a_1^2}+\frac{a_1\nu(m_1+m_2)}
{\sqrt{m_1^2+\nu^2-a_1^2}}\Biggr), \nonumber\\
X_2&=&-\frac{\sqrt{m_1^2+\nu^2-a_1^2}-ia_1}{m_1+i\nu}
-\frac{\l}{2(m_1+i\nu)}\Biggl(2a_1a_2-i\nu(m_1+m_2) \nonumber\\
&&+2ia_2\sqrt{m_1^2+\nu^2-a_1^2}-\frac{a_1\nu(m_1+m_2)}
{\sqrt{m_1^2+\nu^2-a_1^2}}\Biggr), \nonumber\\
X_3&=&\frac{\sqrt{m_2^2+\nu^2-a_2^2}+ia_2}{m_2-i\nu}
+\frac{\l}{2(m_2-i\nu)}\Biggl(2a_1a_2+i\nu(m_1+m_2) \nonumber\\
&&-2ia_1\sqrt{m_2^2+\nu^2-a_2^2}-\frac{a_2\nu(m_1+m_2)}
{\sqrt{m_2^2+\nu^2-a_2^2}}\Biggr), \nonumber\\
X_4&=&-\frac{\sqrt{m_2^2+\nu^2-a_2^2}-ia_2}{m_2-i\nu}
+\frac{\l}{2(m_2-i\nu)}\Biggl(2a_1a_2+i\nu(m_1+m_2) \nonumber\\
&&+2ia_1\sqrt{m_2^2+\nu^2-a_2^2}+\frac{a_2\nu(m_1+m_2)}
{\sqrt{m_2^2+\nu^2-a_2^2}}\Biggr). \label{X_1} \eea

Before passing to $\E$ there still remains to find the expansion
of $r_i=\sqrt{\rho^2+(z-\a_i)^2}$. The calculation of $r_i$ with
the aid of (\ref{alf1}) does not exhibit difficulty and leads to
the expressions \bea
r_1&=&R_1-\l\frac{(z-1)\sqrt{m_1^2+\nu^2-a_1^2}}{R_1}
-\l^2\left(\frac{(m_1^2+\nu^2-a_1^2)(z-1)^2}{2R_1^3}\right.
\nonumber\\
&+&\left.\frac{z(m_1m_2-\nu^2)-m_1^2-m_1m_2+a_1^2}{2R_1}
-\frac{a_1\nu(m_1+m_2)(z-1)}{2R_1\sqrt{m_1^2+\nu^2-a_1^2}}\right),
\nonumber\\ r_2&=&R_1+\l\frac{(z-1)\sqrt{m_1^2+\nu^2-a_1^2}}{R_1}
-\l^2\left(\frac{(m_1^2+\nu^2-a_1^2)(z-1)^2}{2R_1^3}\right.
\nonumber\\
&+&\left.\frac{z(m_1m_2-\nu^2)-m_1^2-m_1m_2+a_1^2}{2R_1}
+\frac{a_1\nu(m_1+m_2)(z-1)}{2R_1\sqrt{m_1^2+\nu^2-a_1^2}}\right),
\nonumber\\ r_3&=&R_2-\l\frac{(z+1)\sqrt{m_2^2+\nu^2-a_2^2}}{R_2}
-\l^2\left(\frac{(m_2^2+\nu^2-a_2^2)(z+1)^2}{2R_2^3}\right.
\nonumber\\
&-&\left.\frac{z(m_1m_2-\nu^2)+m_2^2+m_1m_2-a_2^2}{2R_2}
-\frac{a_2\nu(m_1+m_2)(z+1)}{2R_2\sqrt{m_2^2+\nu^2-a_2^2}}\right),
\nonumber\\ r_4&=&R_2+\l\frac{(z+1)\sqrt{m_2^2+\nu^2-a_2^2}}{R_2}
-\l^2\left(\frac{(m_2^2+\nu^2-a_2^2)(z+1)^2}{2R_2^3}\right.
\nonumber\\
&-&\left.\frac{z(m_1m_2-\nu^2)+m_2^2+m_1m_2-a_2^2}{2R_2}
+\frac{a_2\nu(m_1+m_2)(z+1)}{2R_2\sqrt{m_2^2+\nu^2-a_2^2}}\right),
\label{r_1}\eea where we have introduced
\be
R_1=\sqrt{\rho^2+(z-1)^2}, \quad R_2=\sqrt{\rho^2+(z+1)^2}. \ee

We have realized all the necessary steps for eventually being able
to obtain the form of $\E$ in our approximation scheme. The
substitution of the approximate values of $\a_i$, $X_i$ and $r_i$
into (\ref{DK}) yields, after expanding $\E$ in powers of $\l$,
\bea \E&=&1-2\l\left(\frac{m_1+i\nu}{R_1}+\frac{m_2-i\nu}{R_2}
\right)+2\l^2\left[\left(\frac{m_1+i\nu}{R_1}+\frac{m_2-i\nu}{R_2}
\right)^2\right. \nonumber\\
&&\left.-\frac{ia_1(m_1+i\nu)(z-1)}{R_1^3}
-\frac{ia_2(m_2-i\nu)(z+1)}{R_2^3}\right]. \label{E_1} \eea

For completeness we also give the expanded version of the
potential $\xi$ associates with $\E$: \bea
\xi=\frac{1-\E}{1+\E}&=&
\l\left(\frac{m_1+i\nu}{R_1}+\frac{m_2-i\nu}{R_2}\right)
\nonumber\\ &+&i\l^2\left(\frac{a_1(m_1+i\nu)(z-1)}{R_1^3}
+\frac{a_2(m_2-i\nu)(z+1)}{R_2^3}\right). \label{xi_1} \eea

The knowledge of the potential $\E$ is sufficient for obtaining
the corresponding metric functions $f$, $\omega$, $\gamma$ in
(\ref{Papa}). The function $f$ is simply the real part of $\E$,
hence it has the form \bea
f&=&1-2\l\left(\frac{m_1}{R_1}+\frac{m_2}{R_2}\right)
+2\l^2\left[\left(\frac{m_1}{R_1}+\frac{m_2}{R_2}\right)^2
-\nu^2\left(\frac{1}{R_1}-\frac{1}{R_2}\right)^2 \right.
\nonumber\\ &+&\left.\frac{a_1\nu(z-1)}{R_1^3}
-\frac{a_2\nu(z+1)}{R_2^3}\right]. \label{f_1} \eea

The metric coefficients $\omega$ and $\gamma$ can be found by
integrating equations (\ref{E}) and (\ref{eq_E}), respectively.
The function $\Omega$ entering equations (\ref{E}) is defined by
the imaginary part of $\E$. With this, the integration of
(\ref{E}) gives \bea \omega&=&2\l\nu\left(\frac{z-1}{R_1}
-\frac{z+1}{R_2}\right)-2\l^2\Biggl[\nu(m_1+m_2) \nonumber\\
&+&\rho^2\left(\frac{m_1a_1}{R_1^3}+\frac{m_1a_2}{R_2^3}\right)
-\frac{\nu(m_1+m_2)(z^2+\rho^2-1)}{R_1R_2}\Biggr]. \label{om_1}
\eea

The substitution of (\ref{f_1}), (\ref{om_1}) into equations
(\ref{eq_E}) and the integration of the latter system yields the
form of the remaining function $\gamma$: \bea
\gamma&=&\frac{\l^2}{2}\left[\frac{(m_1m_2-\nu^2)(z^2+\rho^2-1)}
{R_1R_2}-m_1m_2+\nu^2\right. \nonumber\\
&-&\left.\rho^2\left(\frac{m_1^2+\nu^2}{R_1^4}
+\frac{m_2^2+\nu^2}{R_2^4}\right)\right]. \label{g_1} \eea

Mention that the integration constants in the above expressions
for $\omega$ and $\gamma$ are chosen in such a way that both
functions vanish at spatial infinity.

The behavior of $\omega$ and $\gamma$ on the symmetry axis is
crucial for establishing whether equilibrium of two Kerr particles
can be achieved due to balance of the gravitational attraction and
spin--spin repulsion forces. We recall that for two particles to
be in equilibrium, the functions $\omega$ and $\gamma$ should
vanish on the parts of the $z$--axis outside and between the
particles (the particles are situated at the points $z=\pm1$). A
simple inspection of the formulae (\ref{om_1}), (\ref{g_1}) shows
that $\omega=\gamma=0$ when $|z|>1$. By demanding
$\omega=\gamma=0$ on the part $|z|<1$ of the symmetry axis, we
arrive at two balance conditions
\be
-4\nu\l[1+\l(m_1+m_2)]=0 \label{om0_1} \ee (condition of vanishing
$\omega$), and
\be
-\l^2(m_1m_2-\nu^2)=0 \label{g0_1} \ee (condition of vanishing
$\gamma$).

Bearing in mind that the parameter $\l$ is an auxiliary parameter
which can be always put equal to 1 in the final formulae, we see
that the system (\ref{om0_1})--(\ref{g0_1}) has no physically
acceptable solutions since $\nu=0$ satisfying (\ref{om0_1})
implies vanishing of one of the masses $m_1$, $m_2$.

This result about the absence of equilibrium states in the
approximate double--Kerr solution is consistent with our theorem
\cite{MRu1} on the absence of balance of two Kerr black holes with
positive masses in the exact double--Kerr solution.

We turn now to two other approximation schemes within which
equilibrium configurations of two spinning particles are
available.

\vspace{.5cm}

\noindent{\bf 4. Two approximation schemes leading to the balance
of Kerr particles}

\vspace{.5cm}

\noindent The approximation scheme~2 involves two superextreme
Kerr particles. In the exact double--Kerr solution a superextreme
constituent is defined by a cut joining two complex conjugate
$\a$s. As we have seen in the previous section, by supposing the
angular momenta per unit mass to be greater than the masses,
already in the zeroth  approximation we obtained two pairs of
complex conjugate $\a$s as roots of equation (\ref{Sib0_2}).
Hence, in the approximation scheme too a superextreme particle is
represented by a cut, and not by a point on the symmetry axis.
This obviously will lead to the appearance in scheme~2 of four
different square roots $R_i$, $i=\overline{1,4}$, involving
complex quantities, unlike in scheme~1 where we had only two real
distances $R_1$ and $R_2$ defining the location of the particles
on the symmetry axis.

Here we shall consider a slightly modified scheme~2 in which the
parameter $\nu$ will be assigned the order 2 in $\l$ (the case
when $\nu$ is proportional to $\l$, not $\l^2$, does not lead to
equilibrium configurations). The roots of equation
(\ref{alg_cond}) after the substitution $m_1\to\l m_1$, $m_2\to\l
m_2$, $\nu\to\l^2\nu$ into (\ref{axis_DK2}) have been found to
have the following form (technically they are obtainable as in
scheme~1): \bea \a_1=\bar\a_2=1-ia_1+i\l^2\left(
\frac{m_1^2}{2a_1}+\frac{m_1m_2}{a_1+a_2+2i}\right)
+\l^3\frac{\nu(m_1+m_2)}{a_1+a_2+2i}, \nonumber\\
\a_3=\bar\a_4=-1-ia_2+i\l^2\left(
\frac{m_2^2}{2a_2}+\frac{m_1m_2}{a_1+a_2-2i}\right)
-\l^3\frac{\nu(m_1+m_2)}{a_1+a_2-2i}. \label{alf2} \eea

We have found it more advantageous for schemes~2 and 3 to
calculate the quantities $\l X_i$ instead of $X_i$, that
simplifies the subsequent expansion of $\E$. The resulting
expressions for $\l X_i$ are ($\l X_2$ and $\l X_4$ are written
without their quadratic terms in $\l$ because of their complexity)
\bea \l X_1&=&i\l^2\frac{m_1(a_1-a_2+2i)}{2a_1(a_1+a_2+2i)},
\nonumber\\ \l X_2&=&\frac{2ia_1(a_1+a_2-2i)}{m_1(a_1-a_2-2i)}
+\l\frac{2a_1\nu(a_1+a_2-2i)}{m_1^2(a_1-a_2-2i)}, \nonumber\\ \l
X_3&=&-i\l^2\frac{m_2(a_1-a_2+2i)}{2a_2(a_1+a_2-2i)}, \nonumber\\
\l X_4&=&-\frac{2ia_2(a_1+a_2+2i)}{m_2(a_1-a_2-2i)}
+\l\frac{2a_2\nu(a_1+a_2+2i)}{m_2^2(a_1-a_2-2i)}. \label{X_2} \eea

The expansion of $r_i$ leads to the following result: \bea
r_1&=&R_1-i\l^2\frac{z-1+ia_1}{R_1}\left(\frac{m_1^2}{2a_1}
+\frac{m_1m_2}{a_1+a_2+2i}\right), \nonumber\\
r_2&=&R_2+i\l^2\frac{z-1-ia_1}{R_2}\left(\frac{m_1^2}{2a_1}
+\frac{m_1m_2}{a_1+a_2-2i}\right), \nonumber\\
r_3&=&R_3-i\l^2\frac{z+1+ia_2}{R_3}\left(\frac{m_2^2}{2a_2}
+\frac{m_1m_2}{a_1+a_2-2i}\right), \nonumber\\
r_4&=&R_4+i\l^2\frac{z+1-ia_2}{R_4}\left(\frac{m_2^2}{2a_2}
+\frac{m_1m_2}{a_1+a_2+2i}\right), \label{r_2} \eea where
\be
R_1=\bar R_2=\sqrt{\rho^2+(z-1+ia_1)^2}, \quad R_3=\bar
R_4=\sqrt{\rho^2+(z+1+ia_2)^2}. \label{R_2} \ee

From (\ref{DK}) and (\ref{alf2})--(\ref{R_2}) we find the form of
the Ernst potential $\E$: \be \E=1-2\l\left(\frac{m_1}{R_2}
+\frac{m_2}{R_4}\right)+2\l^2\left[\left(\frac{m_1}{R_2}
+\frac{m_2}{R_4}\right)^2-i\nu\left(\frac{1}{R_2}
-\frac{1}{R_4}\right)\right]. \label{E_2} \ee

The corresponding potential $\xi$ is defined by a particularly
simple expression
\be
\xi=\l\left(\frac{m_1}{R_2} +\frac{m_2}{R_4}\right)
-i\nu\l^2\left(\frac{1}{R_2} -\frac{1}{R_4}\right). \label{xi_2}
\ee

Remarkably, the formulae (\ref{E_2}) and (\ref{xi_2}) do not
contain the functions $R_1$ and $R_3$. However, all $R_i$,
$i=\overline{1,4}$, enter the expressions for the metric functions
$f$, $\omega$, $\gamma$ (the procedure of the calculation of these
functions is fully analogous to the one used for scheme~1) \bea
f&=&1-\l\left(\frac{m_1}{R_1}+\frac{m_1}{R_2}+\frac{m_2}{R_3}
+\frac{m_2}{R_4}\right) \nonumber\\
&+&\l^2\left[\frac{m_1^2}{R_1^2}+\frac{m_1^2}{R_2^2}
+\frac{m_2^2}{R_3^2}+\frac{m_2^2}{R_4^2} +\frac{2m_1m_2}{R_1R_3}
+\frac{2m_1m_2}{R_2R_4}\right. \nonumber\\ &+&\left.i\nu\left(
\frac{1}{R_1}-\frac{1}{R_2}-\frac{1}{R_3}+\frac{1}{R_4}
\right)\right], \nonumber\\ \omega&=&i\l\left(
\frac{m_1(z-1+ia_1)}{R_1}+\frac{m_2(z+1+ia_2)}{R_3}-{\rm c.c.}
\right) \nonumber\\ &+&\l^2\left[\frac{m_1^2}{a_1}
+\frac{m_2^2}{a_2}+\frac{2m_1m_2}{a_1+a_2-2i}
+\frac{2m_1m_2}{a_1+a_2+2i}\right. \nonumber\\
&+&\nu\left(\frac{z-1+ia_1}{R_1}-\frac{z+1+ia_2}{R_3} +{\rm c.c.}
\right)-\frac{m_1^2(z^2+\rho^2-2z+a_1^2+1)}{a_1R_1R_2} \nonumber\\
&-&\frac{m_2^2(z^2+\rho^2+2z+a_2^2+1)}{a_2R_3R_4} -2m_1m_2
\nonumber\\
&\times&\left.\left(\frac{z^2+\rho^2+a_1a_2-1-iz(a_1-a_2)
-i(a_1+a_2)}{(a_1+a_2-2i)R_2R_3} +{\rm c.c.}\right)\right],
\nonumber\\ \gamma&=&\l^2\left[\frac{m_1^2}{4a_1^2}
+\frac{m_2^2}{4a_2^2}+\frac{m_1m_2}{(a_1+a_2-2i)^2}
+\frac{m_1m_2}{(a_1+a_2+2i)^2}\right. \nonumber\\
&-&\frac{m_1^2[(z-1)^2+\rho^2+a_1^2]}{4a_1^2R_1R_2}
-\frac{m_2^2[(z+1)^2+\rho^2+a_2^2]}{4a_2^2R_3R_4}-m_1m_2
\nonumber\\ &\times&\left.
\left(\frac{z^2+\rho^2+a_1a_2-1+iz(a_1-a_2)+i(a_1+a_2)}
{(a_1+a_2+2i)^2R_1R_4}+{\rm c.c.}\right)\right] \label{f_2} \eea
(`c.c.' denotes complex conjugation of the terms in brackets).
Interestingly, the constant $\nu$ does not enter the formula for
$\gamma$.

Let us analyze now the possibility of balance of the Kerr
particles in this approximation scheme. By construction, the
functions $\omega$ and $\gamma$ vanish on the part $|z|>1$ of the
symmetry axis. Hence, to achieve the equilibrium of the particles,
we only have to fulfil $\omega=0$ and $\gamma=0$ on the part
$|z|<1$ of the symmetry axis separating the particles. The
condition $\omega=0$ with $R_1=\bar R_2=1-z-ia_1$, $R_3=\bar
R_4=1+z+ia_2$ yields
\be
4\l^2\left(\frac{2m_1m_2(a_1+a_2)}{(a_1+a_2)^2+4}-\nu\right)=0,
\label{om0_2} \ee and the condition $\gamma=0$ leads to
\be
\frac{4m_1m_2\l^2[(a_1+a_2)^2-4]}{[(a_1+a_2)^2+4]^2}=0.
\label{g0_2} \ee

The solution of (\ref{om0_2}) is
\be
\nu=\frac{2m_1m_2(a_1+a_2)}{(a_1+a_2)^2+4}, \label{n_2} \ee and it
fixes the choice of $\nu$ in this approximation scheme. Equation
(\ref{g0_2}) is fulfilled when $a_1+a_2=\pm2$. Recalling that all
the parameters are ``measured'' in the units of $b$, we obtain
after introducing $b$ explicitly:
\be
2b=|a_1+a_2|, \label{b_2} \ee and this is the relation between the
distance at which equilibrium takes place and the angular momenta
per unit mass of the particles.

\underline{\it Scheme~3.} Let us consider now the remaining
approximation scheme. We will show that it also permits
equilibrium configurations, this time between a sub- and a
superextreme Kerr particles, as it occurs in the exact
double--Kerr solution.

The substitutions to be made in the axis data (\ref{axis_DK2}) are
the following: $m_1\to\l m_1$, $m_2\to\l m_2$, $a_1\to\l a_1$,
$\nu\to\l^2\nu$. This means that the particle with subindex~1
represents a subextreme constituent, while the particle with
subindex~2 is a superextreme constituent ($a_2^2>m_2^2$). Mention
that choosing $\nu$ proportional to $\l^2$ is a manner to
introduce the inequality $\nu^2<a_1^2$. As before, we shall
develop this scheme in the second approximation.

Under the above assumptions equation (\ref{alg_cond}) has the
following four roots (the coefficients at $\l^3$ are not given
because of their complexity) \bea \a_1&=&1+\l\sqrt{m_1^2-a_1^2}
+\l^2\frac{m_1m_2}{a_2^2+4}\left(2+\frac{a_1a_2}
{\sqrt{m_1^2-a_1^2}}\right), \nonumber\\
\a_2&=&1-\l\sqrt{m_1^2-a_1^2}
+\l^2\frac{m_1m_2}{a_2^2+4}\left(2-\frac{a_1a_2}
{\sqrt{m_1^2-a_1^2}}\right), \nonumber\\ \a_3&=&\bar\a_4=-1-ia_2
+\l^2\frac{m_2[2m_2+ia_2(2m_1+m_2)]}{2a_2(a_2-2i)}. \label{alf3}
\eea

Then, after expanding $\l X_i$ up to $\l^2$, we get \bea \l
X_1&=&-\l\frac{(a_2-2i)(m_1+ia_1+\sqrt{m_1^2-a_1^2})}
{(a_2+2i)(m_1-ia_1+\sqrt{m_1^2-a_1^2})}+\frac{\l^2}
{m_1^2(a_2+2i)^2} \nonumber\\ &\times&\Biggl[
im_1a_2(2m_1^2-2a_1^2+m_1m_2)\left(1
+\frac{ia_1}{\sqrt{m_1^2-a_1^2}}\right) \nonumber\\
&+&i\nu(a_2^2+4)(\sqrt{m_1^2-a_1^2}+ia_1)\Biggr], \nonumber\\ \l
X_2&=&X_1(\sqrt{m_1^2-a_1^2}\to -\sqrt{m_1^2-a_1^2}), \nonumber\\
\l X_3&=&\l^2\frac{im_2}{2a_2}, \nonumber\\ \l
X_4&=&\frac{2ia_2}{m_2}
+\frac{2a_2\l}{m_2^2}\left(\frac{im_2a_1}{a_2+2i}-\nu\right)
-i\l^2\left(\frac{m_2}{2a_2}+\frac{2a_2\nu^2}{m^3}\right.
\nonumber\\ &-&\left.\frac{2a_2(m_1^2+2a_1^2)}{m_2(a_2+2i)^2}
-\frac{4im_2a_1a_2\nu}{m_2^3(a_2+2i)}\right). \label{X_3} \eea

The expansion of the functions $r_i$ yields \bea r_1&=&R_1
-\l\frac{(z-1)\sqrt{m_1^2-a_1^2}}{R_1}+\frac{\l^2}{2R_1}\left[
(m_1^2-a_1^2)\left(1-\frac{(z-1)^2}{R_1^2}\right)\right.
\nonumber\\ &-&\left.\frac{2m_1m_2(z-1)}{a_2^2+4}\left(2
+\frac{a_1a_2}{\sqrt{m_1^2-a_1^2}}\right)\right], \nonumber\\
r_2&=&r_1(\sqrt{m_1^2-a_1^2}\to -\sqrt{m_1^2-a_1^2}), \nonumber\\
r_3&=&R_3-\l^2\frac{m_2(z+1+ia_2)[2m_2+ia_2(2m_1+m_2)]}
{2a_2(a_2-2i)R_3}, \nonumber\\
r_4&=&R_4-\l^2\frac{m_2(z+1-ia_2)[2m_2-ia_2(2m_1+m_2)]}
{2a_2(a_2+2i)R_4}, \label{r_3} \eea where we have introduced
\be
R_1=\sqrt{\rho^2+(z-1)^2}, \quad R_3=\bar
R_4=\sqrt{\rho^2+(z+1+ia_2)^2}. \label{R_3} \ee

The potentials $\E$ and $\xi$ resulting from the formulae
(\ref{alf3})--(\ref{R_3}) and (\ref{DK}) have the form \bea
\E&=&1-2\l\left(\frac{m_1}{R_1}+\frac{m_2}{R_4}\right)
+2\l^2\left[\left(\frac{m_1}{R_1}+\frac{m_2}{R_4}\right)^2\right.
\nonumber\\ &-&\left. i\nu\left(\frac{1}{R_1}-\frac{1}{R_4}\right)
-\frac{im_1a_1(z-1)}{R_1^3}\right], \nonumber\\
\xi&=&\l\left(\frac{m_1}{R_1}+\frac{m_2}{R_4}\right)
+i\l\left[\frac{m_1a_1(z-1)}{R_1^3}+\nu\left( \frac{1}{R_1}
-\frac{1}{R_4}\right)\right]. \label{E_3} \eea

The set of the metric functions $f$, $\omega$, $\gamma$ defined by
(\ref{E_3}) is given by the expressions \bea f&=&1-\l\left(
\frac{2m_1}{R_1}+\frac{m_2}{R_3}+\frac{m_2}{R_4}\right)
+\l^2\left[\left(\frac{m_1}{R_1}+\frac{m_2}{R_3}\right)^2\right.
\nonumber\\
&+&\left.\left(\frac{m_1}{R_1}+\frac{m_2}{R_4}\right)^2
-i\nu\left(\frac{1}{R_3}-\frac{1}{R_4}\right)\right], \nonumber\\
\omega&=&i\l\left(\frac{m_2(z+1+ia_2)}{R_3}-{\rm c.c.}\right)
+\l^2\left[\frac{m_2[4m_2+a_2^2(4m_1+m_2)]}{a_2(a_2^2+4)}\right.
\nonumber\\ &+&\nu\left(\frac{2(z-1)}{R_1} -\frac{z+1+ia_2}{R_3}
-\frac{z+1-ia_2}{R_4}\right)-\frac{2m_1a_1\rho^2}{R_1^3}
\nonumber\\ &-&\frac{m_2^2[(z+1)^2+\rho^2+a_2^2]}{a_2R_3R_4}
-\frac{2m_1m_2}{R_1}\left(\frac{z^2+\rho^2-1+ia_2(z-1)}
{(a_2-2i)R_3}+{\rm c.c.}\right), \nonumber\\ \gamma&=&\l^2\left[
\frac{m_2^2}{4a_2^2}+\frac{2m_1m_2(a_2^2-4)}{(a_2^2+4)^2}
-\frac{m_1m_2}{R_1}\left(\frac{z^2+\rho^2-1+ia_2(z-1)}
{(a_2-2i)^2R_3}+{\rm c.c.}\right)\right. \nonumber\\
&-&\left.\frac{m_2^2[(z+1)^2+\rho^2+a_2^2]}{4a_2^2R_3R_4}
-\frac{m_1^2\rho^2}{2R_1^4}\right]. \label{f_3} \eea

The metric (\ref{f_3}) is asymptotically flat which means that on
the symmetry axis $\omega=\gamma=0$ when $|z|>1$. The balance of
two Kerr particles then requires $\omega=0$ and $\gamma=0$ for
$|z|<1$. The first of the latter two conditions yields, after
setting $R_1=1-z$, $R_3=1+ia_2+z$, $R_4=1-ia_2+z$,
\be
\frac{4\l^2}{a_2^2+4}[2m_1m_2a_2-\nu(a_2^2+4)]=0, \label{om0_3}
\ee whereas the second condition leads to
\be
\frac{4m_1m_2\l^2(a_2^2-4)}{(a_2^2+4)^2}=0. \label{g0_3} \ee

From (\ref{om0_3}) we get the value of $\nu$ required for the
equilibrium:
\be
\nu=\frac{2m_1m_2a_2}{a_2^2+4}. \label{n_3} \ee

On the other hand, equation (\ref{g0_3}) is satisfied when
$|a_2|=2$. Introducing explicitly the separation parameter $b$, we
finally obtain
\be
2b=|a_2|. \label{b_3} \ee

Formula (\ref{b_3}) defines, for a given value of the angular
momentum per unit mass of the superextreme Kerr constituent, the
coordinate distance at which the equilibrium of two particles
takes place.

Therefore, this approximation scheme~3 too proves to be consistent
with the exact double--Kerr solution.

\vspace{.5cm}

\noindent{\bf 5. The Komar masses and angular momenta in the
approximate solutions}

\vspace{.5cm}

\noindent In this section we shall discuss the relation of the
parameters $m_i$ and $a_i$ of our approximation procedure to the
genuine individual masses $M_i$ and angular momenta $J_i$ of the
particles defined by the Komar integrals \cite{Kom}. For the
metric (\ref{Papa}) in the case of the cylindrical surface of
integration we obtained the formulae \cite{MRS} \bea
4M_i&=&\int\limits_{l_i}^{u_i}\left.\left(\frac{\rho}{f}f_{,\rho}
-\omega\Omega_{,z}\right)\right|_{\rho=\rho_0}dz
+\int\limits_{0}^{\rho_0}\left.\left(\frac{\rho}{f}f_{,z}
+\omega\Omega_{,\rho}\right)\right|_{z=u_i}d\rho \nonumber\\
&-&\int\limits_{0}^{\rho_0}\left.\left(\frac{\rho}{f}f_{,z}
+\omega\Omega_{,\rho}\right)\right|_{z=l_i}d\rho, \nonumber\\
8J_i&=&-\int\limits_{l_i}^{u_i}\left.\left[2\omega
-\frac{2\rho\omega}{f}f_{,\rho}+\left(\frac{\rho^2}{f^2}
+\omega^2\right)\Omega_{,z}\right]\right|_{\rho=\rho_0}dz
\nonumber\\ &&+\int\limits_{0}^{\rho_0}\left.\left[
\frac{2\rho\omega}{f}f_{,z} +\left(\frac{\rho^2}{f^2}
+\omega^2\right)\Omega_{,\rho}\right]\right|_{z=u_i}d\rho
\nonumber\\ &&-\int\limits_{0}^{\rho_0}\left.\left[
\frac{2\rho\omega}{f}f_{,z} +\left(\frac{\rho^2}{f^2}
+\omega^2\right)\Omega_{,\rho}\right]\right|_{z=l_i}d\rho,
\label{Kom} \eea where $u_i$ and $l_i$, $i=1,2$, are the points of
the $z$--axis which are taken as centers of the upper and lower
bases of the cylinder enclosing the $i$th particle, $\rho_0$ being
the radius of the bases. In our calculations we shall assume that
the upper cylinder extends from $z=0$ to plus infinity (i.e.,
$l_1=0$, $u_1=+\infty$), and the other one from minus infinity to
$z=0$ ($l_2=-\infty$, $u_2=0$). The results of the evaluation of
$M_i$ and $J_i$ can be summarized as follows.

\underline{\it Scheme~1.} Although in this scheme there are no
equilibrium configurations and the particles are kept apart by a
supporting strut, it is important for our analysis of the Komar
quantities that the strut be massless in which case we can be sure
that $M_i$ and $J_i$ will not contain any contribution of the
strut. Since the strut is massless when $\omega=0$ on it, we have
to put $\nu=0$ in the formulae (\ref{E_1})--(\ref{g_1}). The
integration with the aid of the formulae (\ref{Kom}) then yields
\be
M_i=\l m_i, \quad J_i=\l^2m_ia_i. \label{mj_1} \ee

Therefore, within the framework of this approximation scheme with
$\nu=0$ the parameters $m_i$ represent the masses of the
particles, and the parameters $a_i$ coincide with the
corresponding angular momenta per unit mass.

\underline{\it Scheme~2.} For the same reasons as in the previous
case, in order to have correct estimations of $M_i$ and $J_i$ it
is sufficient to consider a more general case of particles
supported by a massless strut instead of considering
configurations of balancing particles. So, in the formulae
(\ref{E_2})--(\ref{f_2}) we will take the constant $\nu$ in the
form (\ref{n_2}) to assure $\omega=0$ on the strut, and leave
$a_1$ and $a_2$ as arbitrary parameters. The resulting expressions
for $m_i$ and $J_i$ are
\be
M_i=\l m_i, \quad J_i=\l
m_ia_i+\l^2\frac{4a_im_1m_2}{(a_1+a_2)^2+4}, \label{mj_2} \ee
which means that in this scheme the parameters $a_i$ do not fully
coincide with the Komar angular momenta per unit mass of the
particles.

\underline{\it Scheme~3.} In the third approximation scheme the
massless strut between the particles can be introduced by choosing
$\nu$ in the form (\ref{n_3}). The evaluation of the Komar masses
and angular momenta in (\ref{Kom}) with the aid of
(\ref{E_3})--(\ref{f_3}) leads to the following result:
\be
M_i=\l m_i, \quad J_1=\l^2m_1a_1, \quad J_2=\l
m_2a_2+\l^2\frac{4m_1m_2a_2}{a_2^2+4}. \label{mj_3} \ee

Hence, the parameters $m_i$ in this scheme coincide with the Komar
masses of the particles, and $a_1$ with the angular momentum per
unit mass of the subextreme constituent. The angular momentum per
unit mass of the superextreme constituent is
\be
\frac{J_2}{M_2}=a_2\left(1+\l\frac{4m_1}{a_2^2+4}\right), \ee and
it differs slightly from $a_2$ due to the presence of the second
term in brackets.

\vspace{.5cm}

\noindent{\bf 6. Discussion of the results}

\vspace{.5cm}

\noindent Therefore, we have succeeded in working out correct
approximate analogues of the exact double--Kerr solution by
introducing a small parameter and expanding the Ernst complex
potential in a series of powers of this parameter. It turns out
that the approximation procedure depends drastically on the
qualitative relations between the parameters; it is not unique
since a concrete physical situation dictates the form of the
corresponding approximation scheme. It is worthwhile mentioning
that within all three schemes further variations of the procedure
are possible. For instance, in scheme~1 it is possible to make the
parameter $\nu$ proportional to $\l^2$ or $\l^3$. However, since
these modifications cannot lead in principle to equilibrium
configurations of two subextreme Kerr particles, we did not
include them into this paper. At the same time, one should realize
that by choosing $\nu$ proportional to $\l^3$ he commits himself
to the calculation of the Ernst potential and corresponding metric
functions up to the third order in $\l$ (the system of the balance
conditions in this case takes the form
\be
2\l^3[m_1m_2(a_1+a_2)-2\nu]=0, \quad -m_1m_2\l^2=0, \ee and
implies vanishing of one of the masses.)

A very important qualitative result with which the schemes~2 and 3
provide us is the following: the location of a superextreme
constituent in the approximate solution is not a point on the
symmetry axis but it is a cut determined by two complex conjugate
square roots, say, $R_1=\bar R_2$. One should be aware of this
non-evident result when he wants to develop an approximation
procedure for spinning particles without relating it to a concrete
exact solution.

In view of the above said and of our Section~2 it would be of
interest to compare Bonnor's solution (\ref{sol_B}) with our
approximate solution from scheme~1. The approximation scheme
employed by Bonnor in \cite{Bon3} treats the parameters $m_i$ and
$j_i$ as small parameters of the same magnitude in $\l$. This
means in particular that the quantities $a_i=j_i/m_i$ in his
approach are not small constants and, consequently, both spinning
particles are superextreme because $a_i^2>m_i^2$. The fact that
the superextreme constituents of the solution (\ref{sol_B}) are
placed at the points $\pm b$ of the symmetry axis as if they were
subextreme constituents makes Bonnor's approximation procedure
contradictory from the point of view of the double--Kerr solution.
For the solution (\ref{sol_B}) to represent two subextreme Kerr
particles it is necessary to introduce the small parameters $a_i$
into (\ref{sol_B}) by additionally changing $j_i$ to $\l m_ia_i$
(to make $j_i$ proportional to $\l^2$) and eliminate all the
arising terms of the third order in $\l$. Then after such an
extensive make--up the resulting solution will be a particular
specialization ($\nu=0$) of our approximate solution from scheme~1
(the lack of one arbitrary parameter in the solution (\ref{sol_B})
was discussed in \cite{MRu4}). Mention that in the static limit
the solutions (\ref{sol_B}) and (\ref{f_1})--(\ref{g_1}) coincide
and represent correctly an approximation to the
double--Schwarschild spacetime.

Turning back to our approximate solutions from schemes~2 and 3,
below we would like to demonstrate that the condition (\ref{b_2})
determining the balance of two superextreme Kerr particles and the
condition (\ref{b_3}) determining the balance of a sub- and a
superextreme Kerr constituents agree with the exact results on the
double--Kerr equilibrium problem not only qualitatively but also
quantitatively. Indeed, in our paper \cite{MRu2} we obtained the
general relation for balancing Kerr particles
\be
J\pm(M+s)^2+s(A_1+A_2)=0, \label{r_gen} \ee where
$J=M_1A_1+M_2A_2$, $M=M_1+M_2$, $M_i$ are Komar masses,
$J_i=M_iA_i$ are Komar angular momenta, and $s$ is the coordinate
distance separating the particles ($s=2b$ in the zeroth order in
$\l$).

First let us assume that both constituents in (\ref{r_gen}) are
superextreme. Then, substituting $M_1\to\l M_1$, $M_2\to\l M_2$
into (\ref{r_gen}) and restricting ourselves to only zeroth order
in $\l$ we obtain
\be
\pm s^2+s(A_1+A_2)=0 \quad \Rightarrow \quad s=|A_1+A_2|. \ee This
is exactly the condition (\ref{b_2}).

Assuming now that the first particle is subextreme and the second
one is superextreme, we substitute $M_1\to\l M_1$, $M_2\to\l M_2$,
$A_1\to\l A_1$ into (\ref{r_gen}) and get, after retaining only
the zero--order terms in $\l$:
\be
\pm s^2+sA_2=0 \quad \Rightarrow \quad s=|A_2|, \ee which is
precisely the condition (\ref{b_3}).

\vspace{1.5cm}

\noindent{\bf 7. Conclusions}

\vspace{.5cm}

\noindent The formalism developed in this paper opens
possibilities for the unified treatment of the exact and
approximate solutions of Einstein's equations. It is essentially
based on Sibgatullin's approach to the construction of exact
solutions from the prescribed axis data that permits establishing
very strict and clear connections between the parameters of the
solution and the corresponding physical properties they determine
uniquely through the relativistic multipole moments.

We have constructed approximate analogues of the well--known exact
double--Kerr solution and demonstrated that they permit
equilibrium configurations involving superextreme particles. In
the case of the double--Kerr equilibrium problem the exact
approach looks by far more attractive since the balance conditions
can be solved analytically in a unified form, whereas a specific
approximation procedure has to be worked out for each concrete
type of the Kerr particles. The situation, however, may change in
the case of more complex exact solutions describing the many--body
configurations for which the balance conditions cannot be solved
analytically in general. Then it is anticipated that the
approximate procedure discussed in this paper may become the most
efficient tool for obtaining the general relations for balancing
particles, especially in the presence of the electromagnetic
fields.

\vspace{.5cm}

\noindent{\bf Acknowledgements}

\vspace{.5cm}

\noindent We dedicate this paper to the memory of Nail Sibgatullin
who until his last days was interested in our research and to whom
we are indebted for stimulating and interesting discussions. We
acknowledge partial support from CONACyT of Mexico (project
42729--F) and from DGICyT of Spain (project BFM2000-1322).

\vfill\eject

\newpage

\begin{figure}[htb]
\centerline{\epsfysize=80mm\epsffile{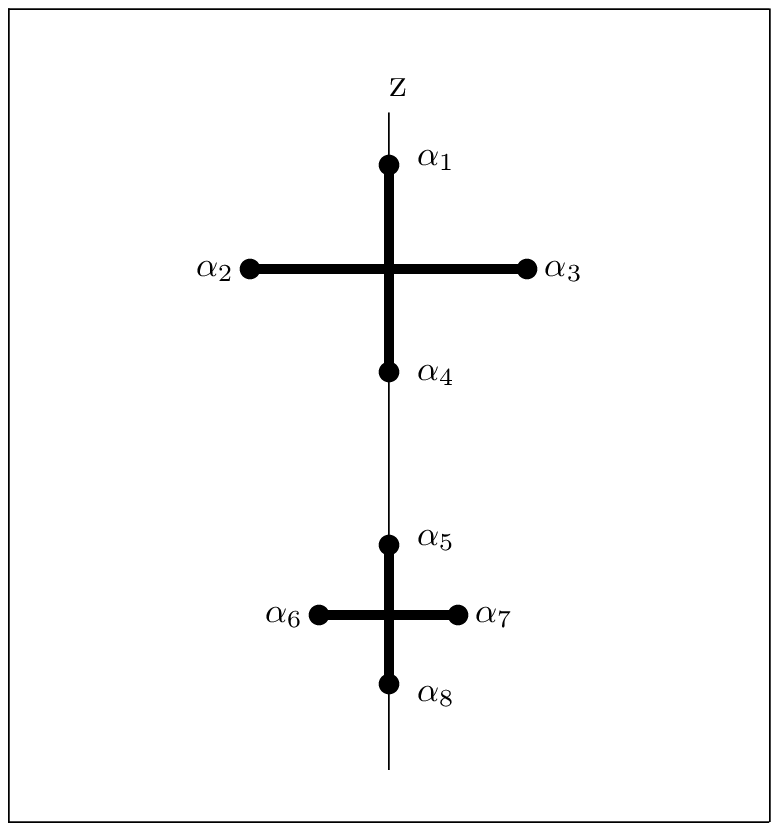}} \caption{Exact
analogue of the approximate solution (4): a specific four--body
system formed by two compound objects.}
\end{figure}

\begin{figure}[htb]
\centerline{\epsfysize=90mm\epsffile{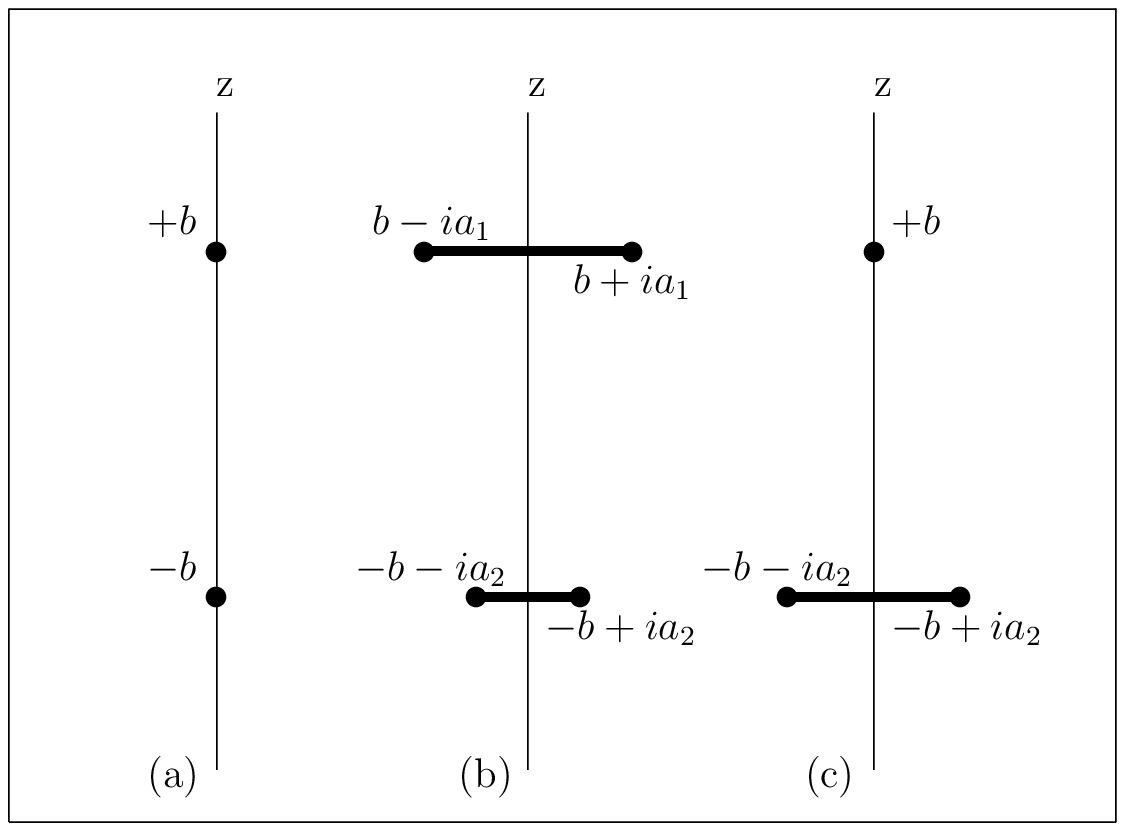}} \caption{Three
principle configurations of two Kerr particles treated by
approximation schemes~1--3.}
\end{figure}


\begin{thebibliography}{99}

\bibitem{Kin} Kinnersley~W \J{1977}{\JMP}{18}{1529}

\bibitem{KCh} Kinnersley~W and Chitre~D~M \J{1978}{\JMP}{19}{2037}

\bibitem{BZa} Belinskii~V~A and Zakharov~V~E \J{1978}{\JETP}{48}{984}

Belinskii~V~A and Zakharov~V~E \J{1979}{\JETP}{50}{1}

\bibitem{Neu} Neugebauer~G \J{1979}{\JPA}{12}{L67}

\bibitem{HEr} Hauser~I and Ernst~F~J \J{1979}{\PRD}{20}{362}

Hauser~I and Ernst~F~J \J{1979}{\PRD}{20}{1783}

\bibitem{HKX} Hoenselaers~C, Kinnersley~W and Xanthopoulos~B~C
\J{1979}{\JMP}{20}{2530}

\bibitem{Sib} Sibgatullin~N~R 1984 {\it Oscillations and Waves in
Strong Gravitational and Electromagnetic Fields} (Moscow: Nauka)
Engl. transl. 1991 (Berlin: Springer)

\bibitem{Bon1} Bonnor~W~B \J{1991}{\PLA}{158}{23}

\bibitem{MPr} Martin~A~W and Pritchett~P~L \J{1968}{\JMP}{9}{593}

\bibitem{Bon2} Bonnor~W~B \J{1993}{\CQG}{10}{2077}

\bibitem{MMS} Manko~O~V, Manko~V~S and Sanabria--G\'omez~J~D
\J{1999}{\GRG}{31}{1539}

\bibitem{BMA} Bret\'on~N, Manko~V~S and Aguilar--S\'anchez~J
\J{1998}{\CQG}{15}{3071}

\bibitem{Wald1} Wald~R \J{1972}{\PRD}{6}{407}

\bibitem{Bon3} Bonnor~W~B \J{2001}{\CQG}{18}{1381}

\bibitem{KNe} Kramer~D and Neugebauer~G \J{1980}{\PLA}{75}{259}

\bibitem{MRS} Manko~V~S, Ruiz~E and Sanabria--G\'omez~J~D
\J{2000}{\CQG}{17}{3881}

\bibitem{MRu1} Manko~V~S and Ruiz~E \J{2001}{\CQG}{18}{L11}

\bibitem{DHo} Dietz~W and Hoenselaers~C \J{1985}{\APNY}{165}{319}

\bibitem{MRu2} Manko~V~S and Ruiz~E \J{2002}{\CQG}{19}{3077}

\bibitem{Ernst} Ernst~F~J \J{1968}{\PR}{167}{1175}

\bibitem{Wald2} Wald~R 1984 {\it General Relativity} (Chicago: The
University of Chicago Press)

\bibitem{FHP} Fodor~G, Hoenselaers~C and Perj\'es~Z
\J{1989}{\JMP}{30}{2252}

\bibitem{Wol} Wolfram~S 1999 {\it The Mathematica Book}, 4th ed.
(Wolfram Media: Cambridge University Press)

\bibitem{MRu3} Manko~V~S and Ruiz~E \J{1998}{\CQG}{15}{2007}

\bibitem{GRG} Hern\'andez--Pastora~J~L, Manko~O~V, Manko~V~S,
Mart\'\i n~J and Ruiz~E \J{2004}{\GRG}{36}{781}

\bibitem{Kom} Komar~A \J{1959}{\PR}{113}{934}

\bibitem{Kerr} Kerr~R~P \J{1963}{\PRL}{11}{237}

\bibitem{MRu4} Manko~V~S and Ruiz~E \J{2003}{\GC}{9}{183}


\end{thebibliography}
\end{document}